# Free-standing cubic gauche nitrogen stable at 760 K under ambient pressure

Yuxuan Xu,[1,2,‡], Guo Chen,[1,2,‡], Fei Du[1,‡], Ming Li[1,2], Liangfei Wu[1], Deyuan Yao[1,2], Junfeng Ding[1,2], Zhi Zeng[1,2], Haiqing Lin[3], and Xianlong Wang[1,2,*]

[1]Key Laboratory of Materials Physics, Institute of Solid State Physics, HFIPS, Chinese Academy of Sciences, Hefei 230031, China;

[2]University of Science and Technology of China, Hefei 230026, China;

[3]Institute for Advanced Study in Physics and School of Physics, Zhejiang University, Hangzhou, 310058, China

[*]Correspondind authors. Emails: xlwang@theory.issp.ac.cn

[‡]These authors contributed equally to this work.

**ABSTRACT**

Cubic gauche nitrogen (cg-N) has received wide attention due to its high energy density and environmental friendliness. However, existing synthesis methods for cg-N predominantly rely on the high-pressure techniques, or the utilization of nanoconfined effects using highly toxic and sensitive sodium azide as precursor, which significantly restrict the practical application of cg-N as high energy density materials (HDEM). Here, based on the first-principles simulations, we find that the adsorption of potassium on the cg-N surface exhibits superior stabilization compared to sodium. Then, we chose the safer potassium azide as raw material for synthesizing cg-N. Through plasma-enhanced chemical vapor deposition treatment, the free-standing cg-N was successfully synthesized without the need of high-pressure and nanoconfined effects. Importantly, it demonstrates excellent thermal stability up to 760 K, and then a rapid and intense thermal decomposition occurs, exhibiting typical behaviors of HDEM thermal decomposition. Our work has significantly promoted the practical application of cg-N as HDEM.

# INTRODUCTION

Since the bond-energy difference between N-N single bond and N≡N triple bond is as high as 794 kJ/mol, a huge amount of energy will be released during the decomposition of polymerized nitrogen with N-N single bond networks into nitrogen gas without producing byproduct of toxic gases [1]. Therefore, polymerized nitrogen is an important potential high-energy-density material (HEDM) and attracts widely attentions [2-4]. The exceptional high-energy and environmentally green features of polymerized nitrogen will make it play an important role in the field of new energy and fuel sectors, etc.

In order to break the strong N≡N triple bond of $N_2$ and give rise to the polymerization, extreme experimental conditions are usually required. High pressure has been proved to be an effective way to synthesized the polymerized nitrogen [5-14]. In 1992, the cubic gauche polynitrogen (cg-N) with a structure similar to diamond was predicted to attain stability at pressures as high as 50 GPa, and it has about five times energy density than that of TNT [2]. Following, many kinds of polymerized nitrogen were theoretically proposed to be stable at pressure higher than 50 GPa [15-43]. In 2004, the cg-N was successfully synthesized using molecular nitrogen at 110 GPa and 2000 K for the first time [10]. Subsequently, various forms of polymeric nitrogen were successfully synthesized, including layered polymeric nitrogen (LP-N at 120 GPa) [11], hexagonal layered polymeric nitrogen (HLP-N at 150 GPa) [12], black phosphorus nitrogen (BP-N at 180 GPa) [13], and Panda nitrogen at 161 GPa [14]. It is worth noting that the phonon spectrum of cg-N, LP-N and HLP-N at atmospheric pressure has no imaginary frequency[12,20], indicating that they can be stabilized to ambient pressure. However, up to now, none of them can be quenched to the ambient pressure, and decompositions generally occurs at pressure of 20~40 GPa [12-13]. Furthermore, the limited synthesis yield also presents a significant obstacle to the practical application of polymeric nitrogen as HEDM.

Therefore, several synthesis strategies have been proposed to produce stable polymeric nitrogen under atmospheric pressure conditions. Chemical pre-pressurization using high-pressure chemical doping was proposed to prepare polymeric nitrogen with a decreased synthesis pressure, and the range of synthesized polymerized nitrogen can be expanded [44-54]. However, the energy densities are decreased notably after doping. Additionally, preserving the polymeric nitrogen at ambient pressure as well as increasing the yield remains a challenge. Apart from chemical doping, researchers also tried polymeric all-nitrogen materials as precursor to synthesized cg-N, which offers the potential to produce cg-N without the need for high-pressure environments. In this approach, the confined cg-N in carbon nanotubes was successfully synthesized at ambient pressure via plasma treatment of sodium azide ($NaN_3$) [55]. But, the process of releasing energy of nanoconfined cg-N can indeed be affected by the presence of carbon nanotubes, and the ratio of cg-N to carbon nanotube is still very low.

Furthermore, understanding the mechanism of cg-N decomposition under low and ambient pressure is crucial for developing efficient methods to product cg-N. Through our investigations of cg-N surfaces under different pressures and temperatures, we found out that the decomposition of cg-N at low pressure is caused by surface instability, and hydrogen saturation can stabilize the cg-N to at least 760 K at ambient pressure due to the saturation of dangling bonds and transferring electrons to the surfaces [58-59]. This finding suggests that hydrogen saturation offers a promising method to enhance the stability and prevent the decomposition of cg-N surfaces. Therefore, in the work about nanoconfined cg-N in carbon nanotubes [55], both carbon nanotubes and sodium may have played a vital role in transferring electrons and stabilizing the cg-N surface. It is also anticipated that the introducing of alkali metals with weaker electronegativity than sodium can further promote the stability of cg-N, such as potassium (K).

In this study, our simulations show that the bonding between potassium and cg-N surface is stronger than that of sodium. Therefore, potassium azide ($KN_3$) is selected as the raw material for synthesizing cg-N by using Plasma Enhanced Chemical Vapor Deposition (PECVD). We successfully synthesized cg-N with a high thermal decomposition temperature of 760 K at ambient pressure. Furthermore, our method has several advantages: I. Without carbon nanotubes for nanoconfinement, the cg-N production can be easily scaled up, and the energy release will be more efficient as there is no additional energy required to break the confinement of the nanostructures; II. Unlike $NaN_3$, which is highly toxic, sensitive, and explosive, $KN_3$ is comparatively safer to handle, which is more readily accessible for research and industrial applications.

**METHOD**

**Calculated part**

The first-principles calculations of the geometrical relaxation, adsorption energy and electronic properties were performed using the Vienna ab initio approximation package (VASP) [58-59] based on the density functional theory with a plane-wave basis set employing projector augmented wave (PAW) method. The generalized gradient approximation (GGA) [60-61] parameterized by the standard Perdew-Burke-Ernzerhof (PBE) [62] exchange-correlation functional was used. An energy cutoff of 520 eV was applied in all calculations, and the convergence criteria for energy and force were set to $1 \times 10^{-6}$ eV and 0.001 eV/Å, respectively. The k-point grid was generated using the Monkhorst-Pack scheme with a k-grid spacing of 0.25 Å$^{-1}$, ensuring a K-point density that converged the energy within 0.5 meV/atom. Post-processing was performed using MULTIWFN [63-64]. Adsorption energies are calculated using the equation of $E_{abs} = (E_{total} - nE_i - E_{pri})/n$, where $E_{ads}$, $E_{pri}$, $E_i$, and $n$ is the total energy of surfaces with adsorption, the energy of pristine surface, the energy of isolated atom calculated from the total energy of the Na/K crystal, and the number of adsorbates, respectively. Negative adsorption energy means that the adsorption process is energy-favorable.

**Experimental part**

Experiments: 200 mg $KN_3$ (purchased from Meryer Inc) was placed in a crucible and then was put into the PECVD furnace for cg-N synthesis. The argon (Ar) and nitrogen (N$_2$) gases are commonly used as carrier gases and plasma-forming gases in PECVD processes, and the flow rate of both Ar and N$_2$ were maintained at 15 standard cubic centimeters per minute (sccm). The power input to the PECVD system was set to ~130 W, which is used to generate and sustain the plasma within the reaction chamber. The reaction temperature and time were set to about 300°C and 3h, respectively, which aids the decomposition and reaction of the KN$_3$ precursor.

Characterization: Measurements of Raman spectra were carried out with a Invia Qontor Raman spectrometer (Renishaw, UK) in range of 100~3000 cm$^{-1}$ with 532 nm laser excitation at a spectral resolution of 1 cm$^{-1}$. XRD measurements were conducted with a TCU 2000 Temperature Contral Unit X-ray Diffractometer (Anton Paar Instrument Corporation, AT) in the range from 20 to 90°. Fourier transform infrared (FTIR) spectroscopy measurements were performed using a NEXUS 670 instrument (Thermo Nicolet, USA) with a range of 400~4000 cm$^{-1}$. The thermal performance of samples were characterized by thermogravimetry-differential scanning calorimetry (TG-DSC) (TGA/DSC3+, METTLER, CH). TG-DSC were measured at a ramp rate of 10°C/min under nitrogen atmosphere.

**RESULTS AND DISCUSSION**

Our previous results show that the pristine (111) surface of cg-N is stable at the ambient condition [56], and it is a good model to investigate the adsorption behaviors of alkai metals. The supercell of (111) surface containing 192 atoms is shown in the Fig. 1(a) after structural relaxation. The structures of one, two, and three Na/K atoms adsorbed onto the surface are shown in Fig. 1(b), and we can find that the adsorption energies of both Na and K atoms are negative, indicating that they tend to adsorb on the (111) surface. Notably, the adsorption energy of one K atom (-4.319 eV) is comparatively lower than that of one Na atom adsorption (-4.072 eV), indicating a preference for stronger adsorption. The same phenomenon occurs in the adsorptions of two and three Na/K atoms, which indicates a persistent trend wherein K atoms exhibits a consistently lower adsorption energy compared to Na atoms across all cases studied. In other words, K atoms can easily adsorb onto the surface and adsorb more firmly than Na atoms.

The electron density difference (EDD) and local integration is calculated for illustrating the charge transfer behaviors of (111)-Na and (111)-K, and the results are shown in Fig. 2, confirming the charge transfer from Na/K atoms into the (111)-surface. Furthermore, local integration quantitatively reveals that K atom imparts a higher degree of charge to the surface in comparison to Na atom. Please note, the conclusion drawn from our previous work is that transferring more electrons to the surface can further promote surface stability [56]. Therefore, compared to Na atoms, adsorption of K atom has an advantage in increasing the stability of cg-N, due to its ability to adsorb more firmly and transfer more electrons.

Following, the effects of K atom adsorption on the nitrogen bonds will be discussed. For pristine surface, the presence of dangling bonds will make the N-N bond length on the surface shorter. As shown in Fig. 3, the N-N bond length on the (111)-surface is 1.383 Å, which is notably shorter than that inside the crystal (~1.410 Å). After one, two, and three K atoms adsorption, the surface N-N bond length monotonically increase to 1.387 Å, 1.391 Å, and 1.395 Å, respectively, reverting to that of the crystal. The bond lengths at a distance of 4.5 Å from the surface can also be slightly affected by the adsorption of K atoms. The results indicate that the appearance of K atoms can indeed have notable impact on the bond length of the surface.

The distribution behaviors of K atoms on the (111)-surface are investigated by putting two K atoms at different distances on the surface and then conducting structural relaxation. The initial and final structures are presented in Fig. 4, three initial K-K distances (1.867 Å, 2.614 Å, and 4.107 Å) increases with structural relaxation, ultimately reaching the same value of 6.254 Å, resulting in a homogeneously hexagonal distribution. This distribution behavior further indicates that adsorption of K atoms is very suitable for stabilizing the surfaces of cg-N. Therefore, in our experiments, we used $KN_3$ as the precursor to synthesis cg-N at ambient pressure.

Raman spectra obtained from unreacted and plasma-reacted $KN_3$ are shown in Fig. 5(a). The characteristic peaks associated with unreacted $KN_3$ at 1276 $cm^{-1}$ and 1347 $cm^{-1}$ assigned to the $2\nu_2$ and $\nu_1$ modes of azide ion, respectively. After treating $KN_3$ with plasma reaction, a new peak at 640 $cm^{-1}$ was observed. FTIR spectra of unreacted and plasma-reacted $KN_3$ are depicted in Fig. 5(b). Without the plasma treatment, only the peaks corresponding to the bending $\nu_2$ mode at 640 $cm^{-1}$ and the antisymmetric stretching $\nu_3$ mode at 2100 $cm^{-1}$ of azide ions can be observed. The peak at 1432 $cm^{-1}$ is assigned to a modified stretching mode of the azide ions in phase I [55]. After the plasma treatment of $KN_3$, the synthesized samples exhibited new low-intensity vibrational peaks at 883 $cm^{-1}$. By comparing the above results with previous experimental reports of CNT-limited domains [55] and theoretical results [64], it can be proved that the new Raman peak at 640 $cm^{-1}$ is the characteristic peak of cg-N, and the peak at 883 $cm^{-1}$ with lower intensity in FTIR spectra can be clearly assigned to the T(TO) symmetry vibration of cg-N at ambient temperature and pressure, which further supports the conclusion that the obtained sample is cg-N. In order to investigate the stability of cg-N, the samples were stored in sealed bags and tested the Raman spectral after 60 days, respectively. As can be seen from Fig. 5(a), cg-N could be stabilized at ambient temperature and pressure for at least 2 months. The degradation of cg-N may be due to the relatively active state of K atoms at the cg-N surface, which leads to a gradual shift in the chemical state of cg-N.

The XRD spectra of $KN_3$ before and after the plasma treatment is shown in the Fig. 6. Upon plasma reaction, the XRD spectra of the samples exhibited new peaks near 80°, which were not present in the unreacted $KN_3$. The peak near 80° aligns with our simulated XRD pattern, and the remaining peaks from the calculated results also match those observed in the sample. This strong agreement further proving that the synthesized product is indeed cg-N.

As shown in Fig. 7, TG-DSC was employed to analyze the samples of $KN_3$ before and after plasma treatment, and the results were obtained using a heating rate of 10 K/min. From the DSC curve in Fig. 7(a), it is evident that the raw material $KN_3$ exhibits a melting and heat absorption peak in the temperature range of 325~350°C, which indicate that $KN_3$ undergoes a phase transition from the solid state to the liquid state. An exothermic decomposition peak was observed at 462°C. By contrast, the $KN_3$ after being treated with plasma possess a similar melting heat absorption peak between 325~350°C. Based on this observation, it can be inferred that the plasma treatment has limited penetration depth, and the interior of the cg-N sample remains predominantly composed of $KN_3$. The exothermic decomposition peaks of the sample were observed at 488°C and 507°C. Based on the comparison, it can be inferred that the peak at 488°C corresponds to the decomposition of cg-N, while the peak at 507°C exhibited a similar peak shape with that of $KN_3$ at 462°C, but with less heat released. The elevated decomposition temperature of $KN_3$ in the PECVD treated sample may be caused by the presence of cg-N on the outer layer.

As observed from the TG curves in Fig. 7b, the weight loss during decomposition of cg-N at 488°C was 0.7%. In contrast, the weight loss of $KN_3$ at 507°C was close to 90%. However, comparing the DSC curves, the heat flow peak of the cg-N decomposition at 488°C is highly variable, with an intensity of about half that of $KN_3$ at 507°C. Therefore, the TG-DSC results suggests cg-N is a highly energy-containing substance that can undergo a pronounced exothermic effect in a short period of time.

**CONCLUSION**

In conclusion, first-principles calculations were employed to compare the adsorption energies and EDD for both Na and K on the (111a)-surface, which allowed for a quantitative assessment of the adsorption behavior and exploration of preferential adsorption sites for K atoms. Experiments were conducted to synthesize cg-N based on the PECVD using $KN_3$ as a precursor.

The findings show that K exhibits superior adsorption stability when interfaced with the cg-N surface. The successful synthesis of cg-N was confirmed through the utilization of various characterization techniques, including Raman spectroscopy,

FTIR, XRD, TG-DSC. Importantly, the synthesized cg-N demonstrated remarkable stability for at least two months under atmospheric conditions. The TG-DSC results demonstrate that cg-N can be stabled at a temperature of 760 K and have a good capacity of exothermic.

The results of this study highlight the advantage of using $KN_3$ as a precursor for the synthesis of cg-N, which not only demonstrates the feasibility of producing cg-N but also holds promise for inspiring future developments in the field of high-energy-density materials. To improve the productivity and stability of cg-N, in-depth investigations on the growth mechanism of cg-N and optimization of the synthesis process are our following task.

**FUNDING**

This work is supported by the National Natural Science Foundation of China (NSFC) under Grant of U2030114 and 12088101, and CASHIPS Director's Fund (Grant No. YZJJ202207-CX, YZJJ202308-TS, YZJJ-GGZX-2022-01). The calculations were partly performed in Center for Computational Science of CASHIPS, the ScGrid of Supercomputing Center and Computer Network Information Center of Chinese Academy of Sciences, and the Hefei Advanced Computing Center.

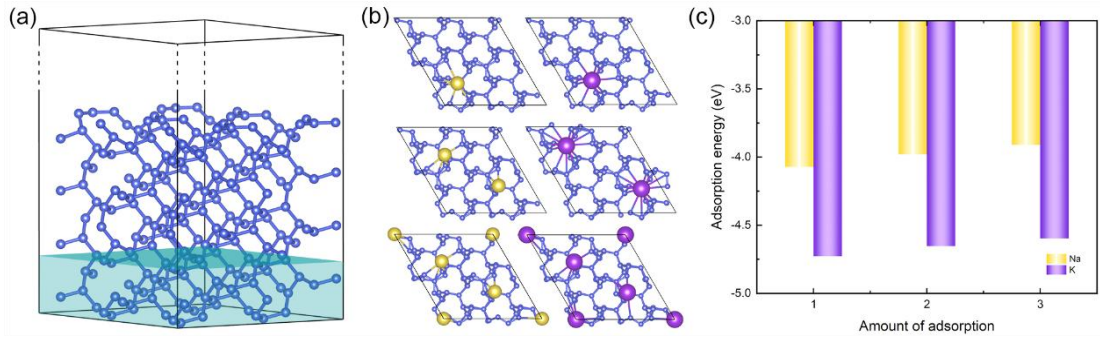

**Figure 1.** Configurations and Na/K atoms' adsorption energy for (111) surfaces. (a) Configuration of relaxed (111)-surface. (b) The top view of different amounts of adsorption for (111)-Na and (111)-K. (c) Variation in adsorption energy for Na and K on the (111)-surface. The atoms located in the blue shadow area are fixed. To clearly show the atomic structure of surfaces, the vacuum between surfaces is not fully presented, and it is indicated by the dotted lines. Blue, yellow, purple balls represent N, Na, and K atoms, respectively.

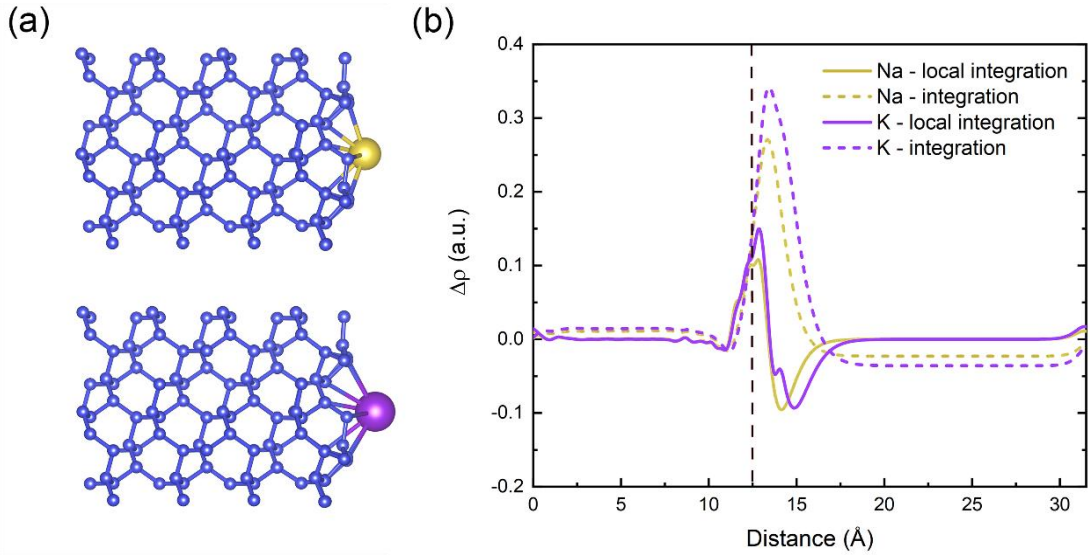

**Figure 2.** Energy density difference (EDD) for (111)-Na and (111)-K surfaces. (a) Configuration of (111)-Na and (111)-K for EDD model. (b) Local integral curves (solid lines) and full integral curves (dotted lines) for EDD. The horizontal coordinates represent the distance along the z axis. The local integral curve illustrates the electron gain (positive values) and loss (negative values) along the z axis, and the integral curve is the integration of the local integral curve. The black dashed line represents the position of the surface. The yellow, and purple atoms represent Na and K atoms, respectively.

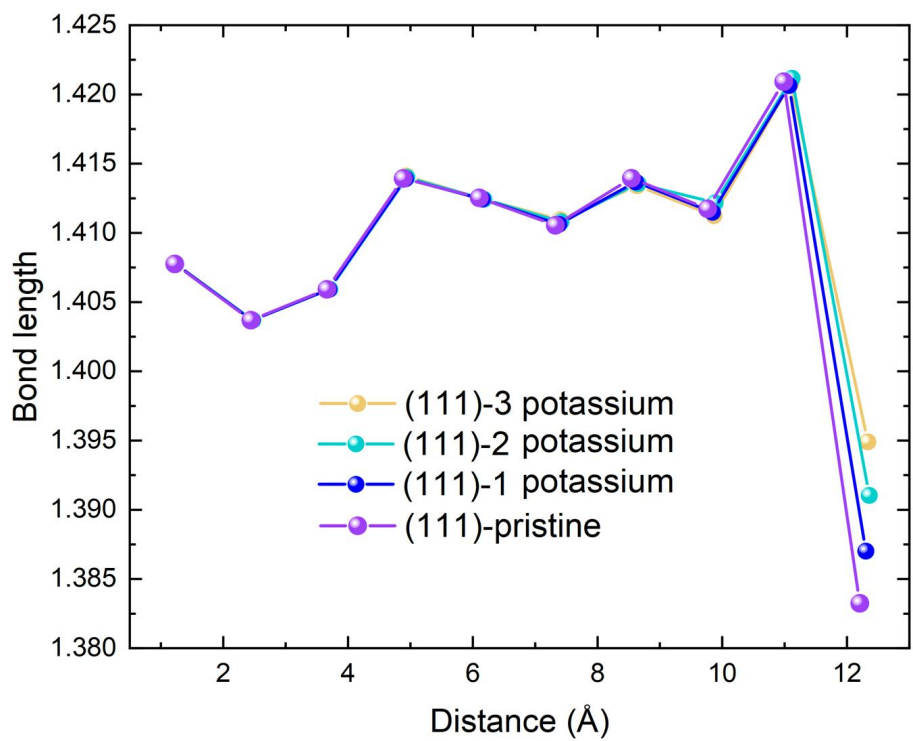

**Figure 3.** Variation of N-N bond length with different numbers of K atoms adsorbed on the (111)-surface. The horizontal coordinates represent the distance along the z axis of the structure, with points indicating the average N-N bond length for the respective positions.

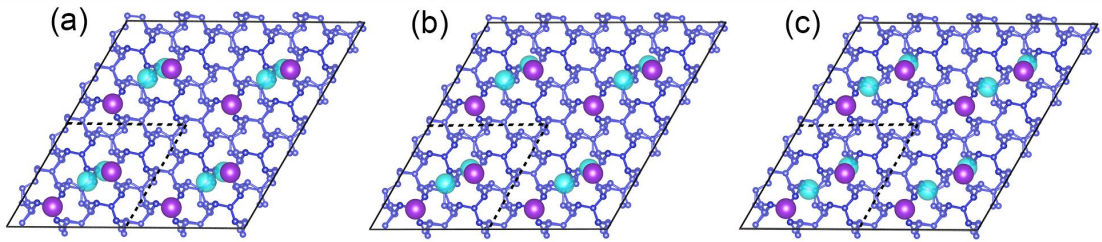

**Figure 4.** Structures of K atoms on the 111 surfaces with initial K-K distance of 1.867 Å (a), 2.614 Å (b), and 4.107 Å (c). The green atoms represent the initial positions, while the purple atoms depict the final positions. The dashed box indicates the supercell used for the calculations.

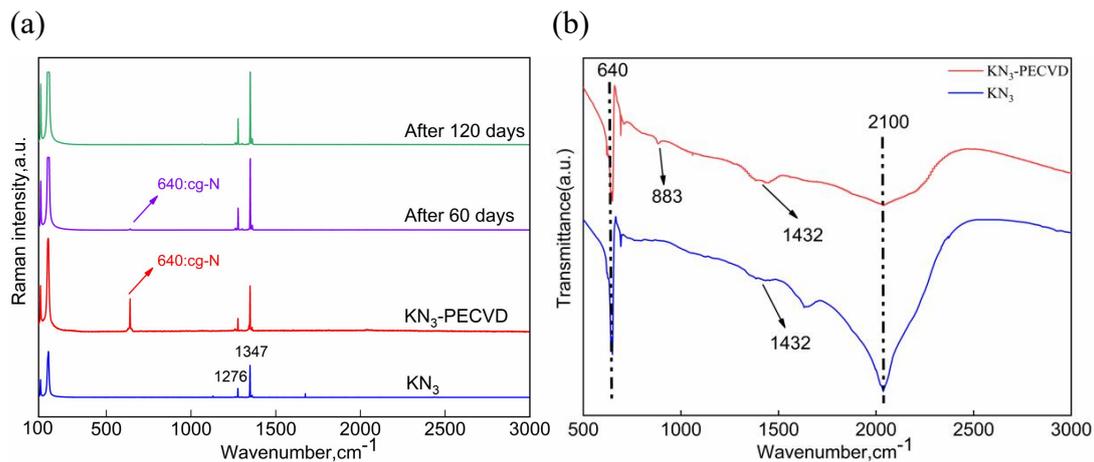

**Figure 5.** (a) Raman spectra of the samples before and after plasma reaction. $KN_3$ has two Raman characteristic peaks at 1267 cm$^{-1}$ and 1360 cm$^{-1}$ assigned to the 2ν2 and ν1 modes of azide ion. cg-N has one Raman characteristic peak at 640 cm$^{-1}$. (b) FTIR spectra of the $KN_3$ before and after reaction with plasma. The characteristic peaks of $KN_3$ are the bending ν2 mode at 640 cm$^{-1}$ and the antisymmetric stretching ν3 mode at 2100 cm$^{-1}$ of azide ions. The cg-N has the characteristic peak at 883 cm$^{-1}$.

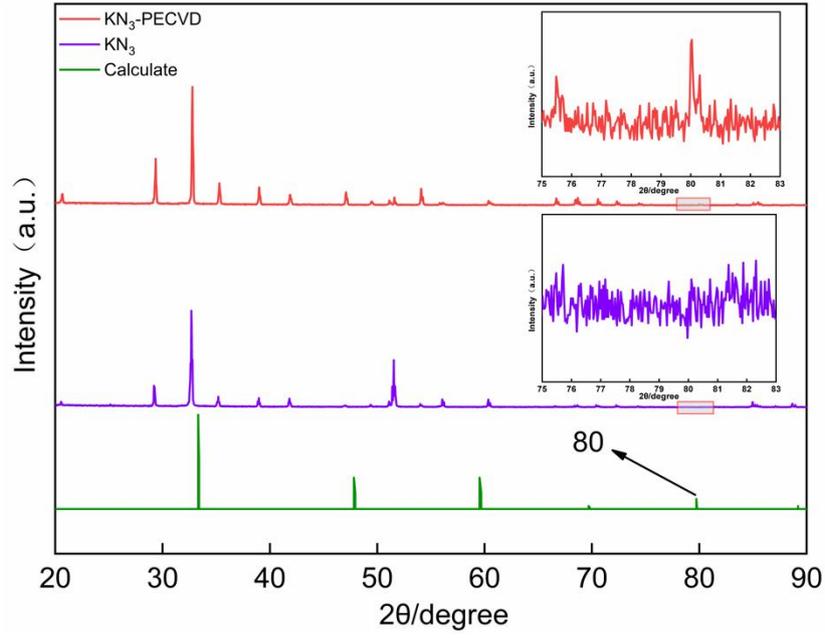

**Figure 6.** XRD patterns of KN$_3$ before and after plasma reaction and the simulated result of cg-N. The new appeared peak at 80° after plasma treatment can match with the peak of theoretical simulations.

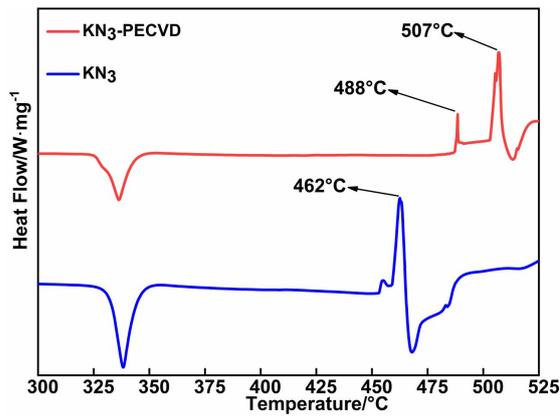 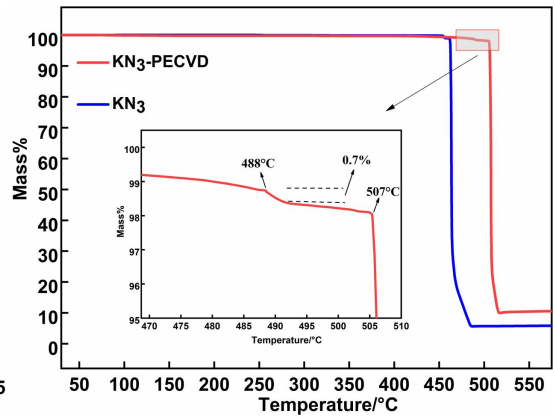

**Figure 7.** (a) DSC and (b) TG curves of KN$_3$ before and after plasma treatment. KN$_3$ has a melting and heat absorption peak between 325~350°C. The exothermic decomposition of the cg-N peaks is observed at 488°C. There has a 0.7% weight loss at 488°C.